\documentclass{eas}
\usepackage{graphicx}
\begin{document}

\runningtitle{RAVE as a Gaia precursor: what to expect from the Gaia RVS?}

\newcommand{\kms}{km~s$^{-1}$}
\newcommand{\teff}{T$_{\rm eff}$}
\newcommand{\logg}{$\log~g$}
\newcommand{\feh}{$\rm [Fe/H]$}
\newcommand{\meta}{${\rm [M/H]}$}

%%-----------------------------
%%      the top matter
%%-----------------------------
\title{RAVE as a Gaia precursor: what to expect from the Gaia RVS?} 
\author{Matthias Steinmetz for the RAVE collaboration}\address{Leibniz-Institut f\"ur Astrophysik Potsdam (AIP), An der Sternwarte 16, 14482 Potsdam, Germany}

\begin{abstract}
The Radial Velocity Experiment (RAVE) is a large wide-field spectroscopic stellar survey of the Milky Way. Over the period 2003-2013, 574,630 spectra for 483,330 stars have been amassed at a resolution of R=7500 in the Ca-triplet region of 8410-8795\AA. Wavelength coverage and resolution are thus comparable to that anticipated from the Gaia RVS. Derived data products of RAVE include radial velocities, stellar parameters, chemicals abundances for Mg, Al, Si, Ca, Ti, Fe, and Ni, and absorption measures based on the diffuse interstellar bands (DIB) at 8620\AA. Since more than 290000 RAVE targets are drawn from the Tycho-2 catalogue, RAVE will be an interesting prototype for the anticipated full Gaia data releases, in particular when combined with the early Gaia data releases, which contain astrometry but not yet stellar parameters and abundances.
\end{abstract}
\maketitle
%%-----------------------------
%%      your text
%%-----------------------------
\section{Introduction}
Despite the fact that the Galaxy is one unique system, understanding its formation holds important clues to study the broader context of galaxy formation.  Wide field spectroscopic surveys play a particularly important role in analysis of the Milky Way: Spectroscopy enables us to measure the radial velocity, which in turn allows us to study the details of Galactic dynamics. Spectroscopy also permits to measure the abundance of chemical elements in a stellar atmosphere which holds important clues on the initial chemical composition and its subsequent metal enrichment. Despite this importance, ten years ago wide-field spectroscopic surveys of the Milky Way was still limited to the Geneva Copenhagen survey (CGS, Nordstr\"om et al, 2004), which only covered a sphere of about 100~pc radius around the sun (the so-called Hipparcos sphere).

The situation has fundamentally changed over the past decade, with several wide-field spectroscopic surveys underway: SDSS-SEGUE and RAVE being completed in terms of data taking, LAMOST, APOGEE and HERMES well underway, and some massive campaigns such as 4MOST, WEAVE and DESI in the making, each of the latter delivering spectroscopic data for some 10 million stars. The Gaia mission will also not only provide exquisit distances and proper motions for up to 1 billion stars, it also will deliver spectra for some 100 million stars. The RAVE survey can play a particular role in the preparation of the Gaia era, as the spectral range, signal to noise and spectral resolution of RAVE spectra is similar to those expected from the Gaia-RVS\footnote{For visualization of the volume covered by some of these surveys, see http://www.rave-survey.org/project/gallery/movies/}.

\section{RAVE survey description}
RAVE\footnote{http://www.rave-survey.org} began observations in 2003, and since then has released four{  data releases (noted DR hereafter)}: DR1 in 2006 (Steinmetz \etal\ 2006), DR2 in 2008 (Zwitter \etal\ 2008), DR3 in 2011 (Siebert \etal\ 2011), and DR4 in 2013 (Kordopatis \etal\ 2013a).  RAVE is a magnitude-limited survey of stars randomly selected in the southern celestial hemisphere. The original design was to only observe stars in the interval $9<I<12$ but owing to the IR based selection function, some stars that are brighter and fainter can be found. The spectra are obtained from the 6dF facility on the 1.2m Australian Astronomical Observatory's UK-Schmidt telescope in Siding Spring, Australia, where three field plates with up to 150 robotically positioned fibres are used in turn.
The effective resolution of RAVE is $R=\lambda/\Delta \lambda \sim7500$ and the wavelength range coverage is around the infrared  ionised Calcium triplet (IR Ca{\footnotesize II}, $\lambda \lambda 8410-8795$\AA), one of the widely used wavelength ranges for Galactic archaeology. It is also the wavelength range in which the Gaia RVS is operating. RAVE overall has amassed 574 630 spectra for 483 330 unique stars,  425 561 of these targets are included in DR4. 

\begin{figure}
\centering
\includegraphics[width=0.9\textwidth]{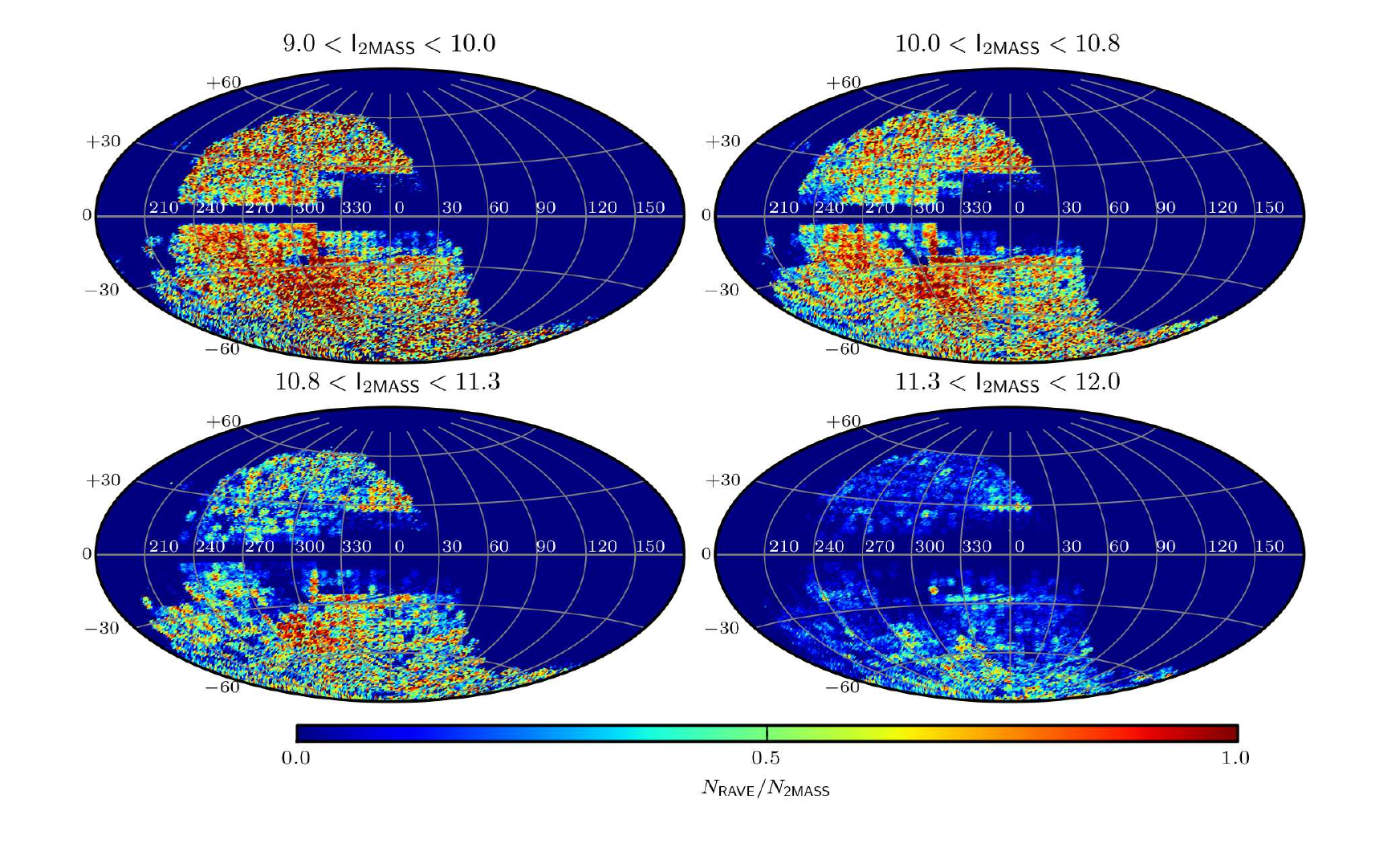} 
\caption{Aitoff projection (Galactic coordinates) of the completeness of the stars in the $I_{\rm 2MASS}$ band for which radial velocity measurements are available. Each panel shows a different magnitude bin. Grey-scale coding represents the ratio of RAVE observations to 2MASS stars.  (from DR4) }
\label{Fig:Completeness_maps}
\end{figure}

DR4's completeness with respect to 2MASS is shown for various magnitude bins in the Aitoff projections shown in Figure~\ref{Fig:Completeness_maps}.   Each 6dF field set-up only selects  targets from one of the following four magnitude bins: $9.0 < I < 10.0$ ($8.0 < I < 10.0$ for the new 2MASS input catalogue), $10.0 < I < 10.8$,  $10.8 < I < 11.3$ and $11.3 < I < 12$ mag.   This minimizes the magnitude range within any one 6dF field set-up to be within a bin, meaning exposure times can be scaled more appropriately for all the targets in the same field.  Each field set-up is a random selection of unobserved targets within these bins (apart from designed repeat observations).  Any spectrum within a 6dF field set-up can be adjacent to any other in the same set-up on the CCD but the bins limit the magnitude difference, which also minimizes fibre cross-talk. Figure~\ref{Fig:Completeness_maps} also shows that while for the brighter magnitudes bins, high rates of completeness have been achieved (with some individual fields reaching close to 100\% compared to 2MASS), RAVE has not exhausted the input catalogue for fainter targets, in particular for $I>11.3$.

\section{RAVE data products}
\subsection{Radial velocities}
\begin{figure}[btp]
\includegraphics[height=0.3\textwidth]{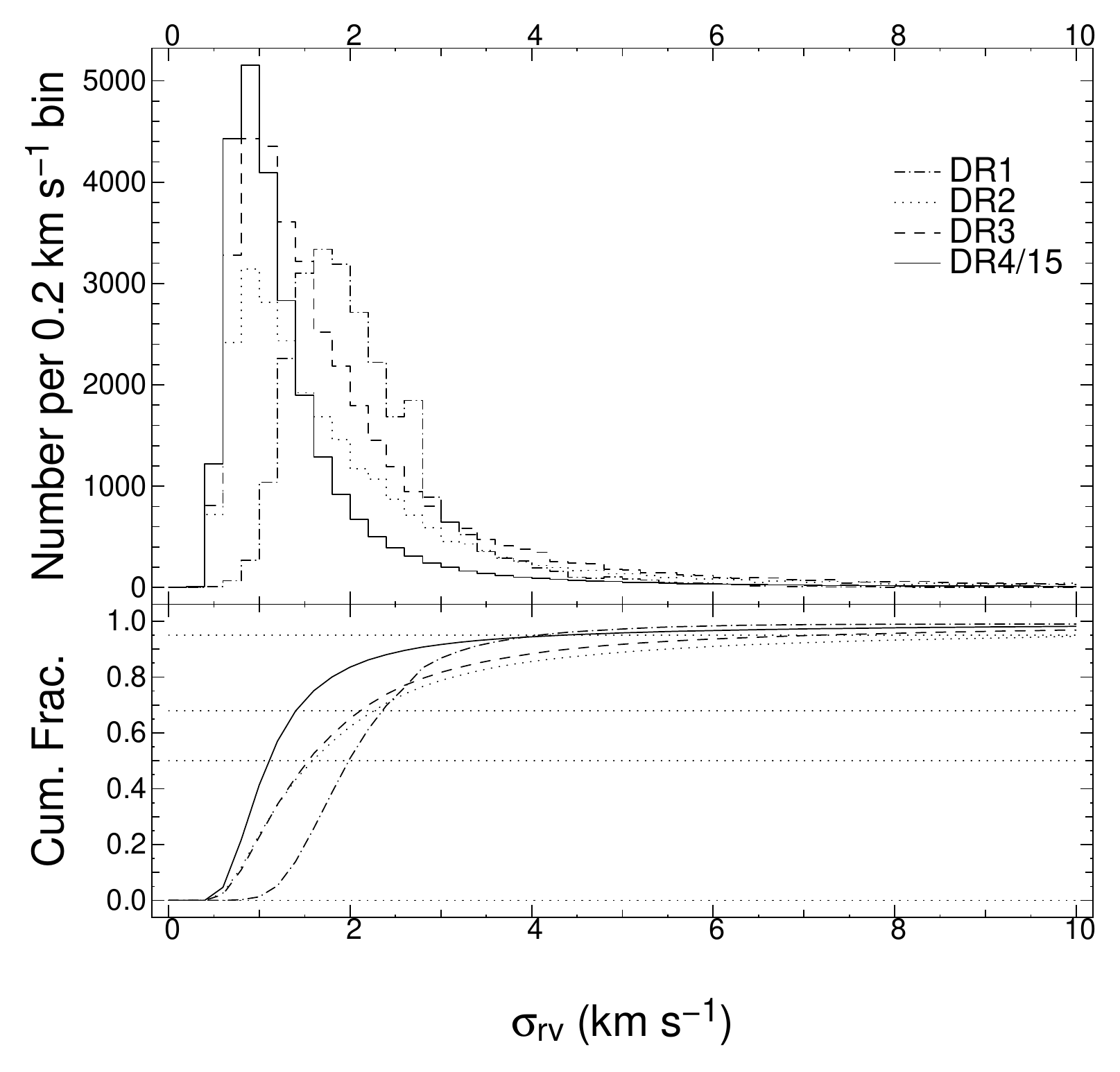}
\includegraphics[height=0.29\textwidth]{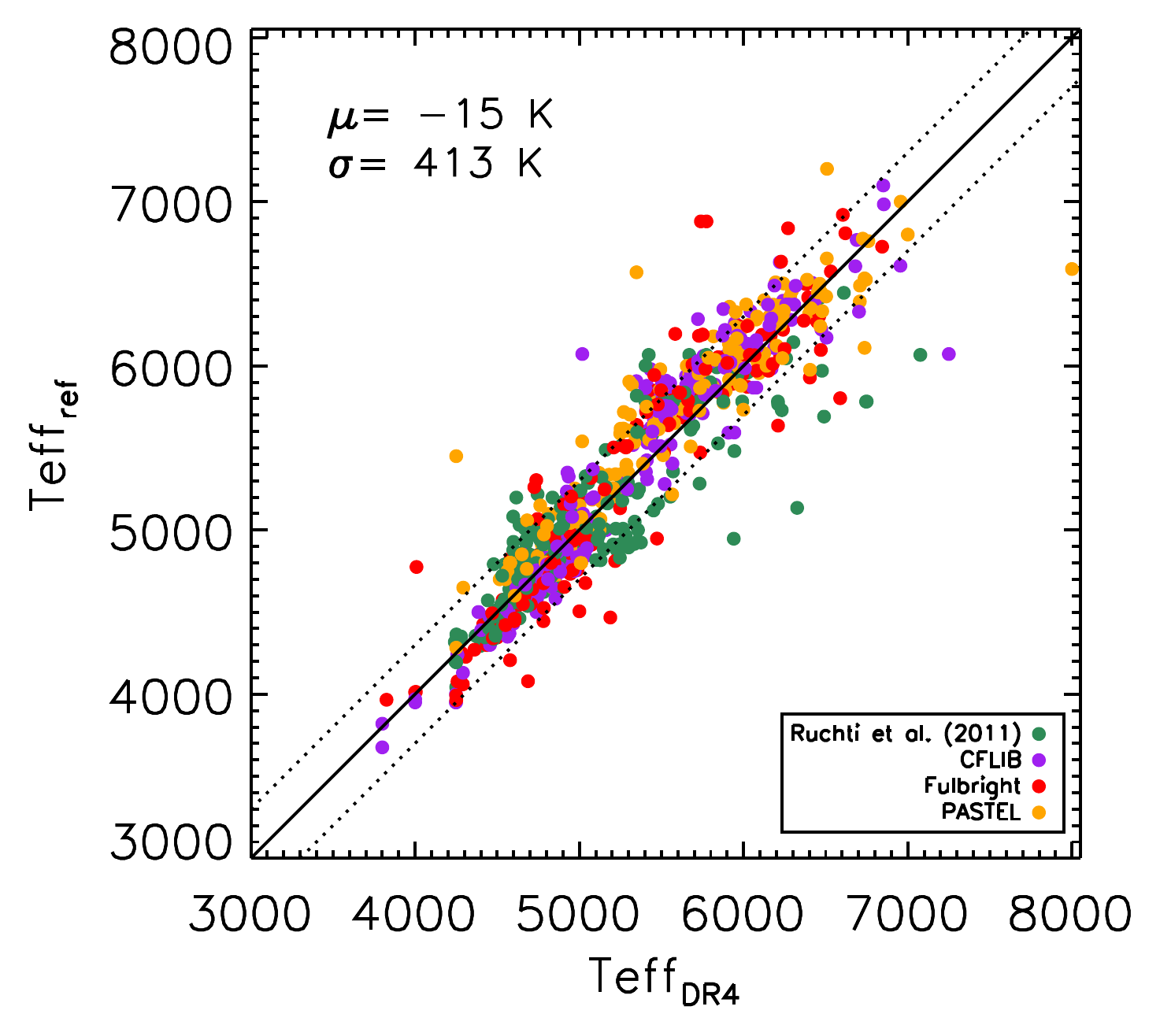}
\includegraphics[height=0.29\textwidth]{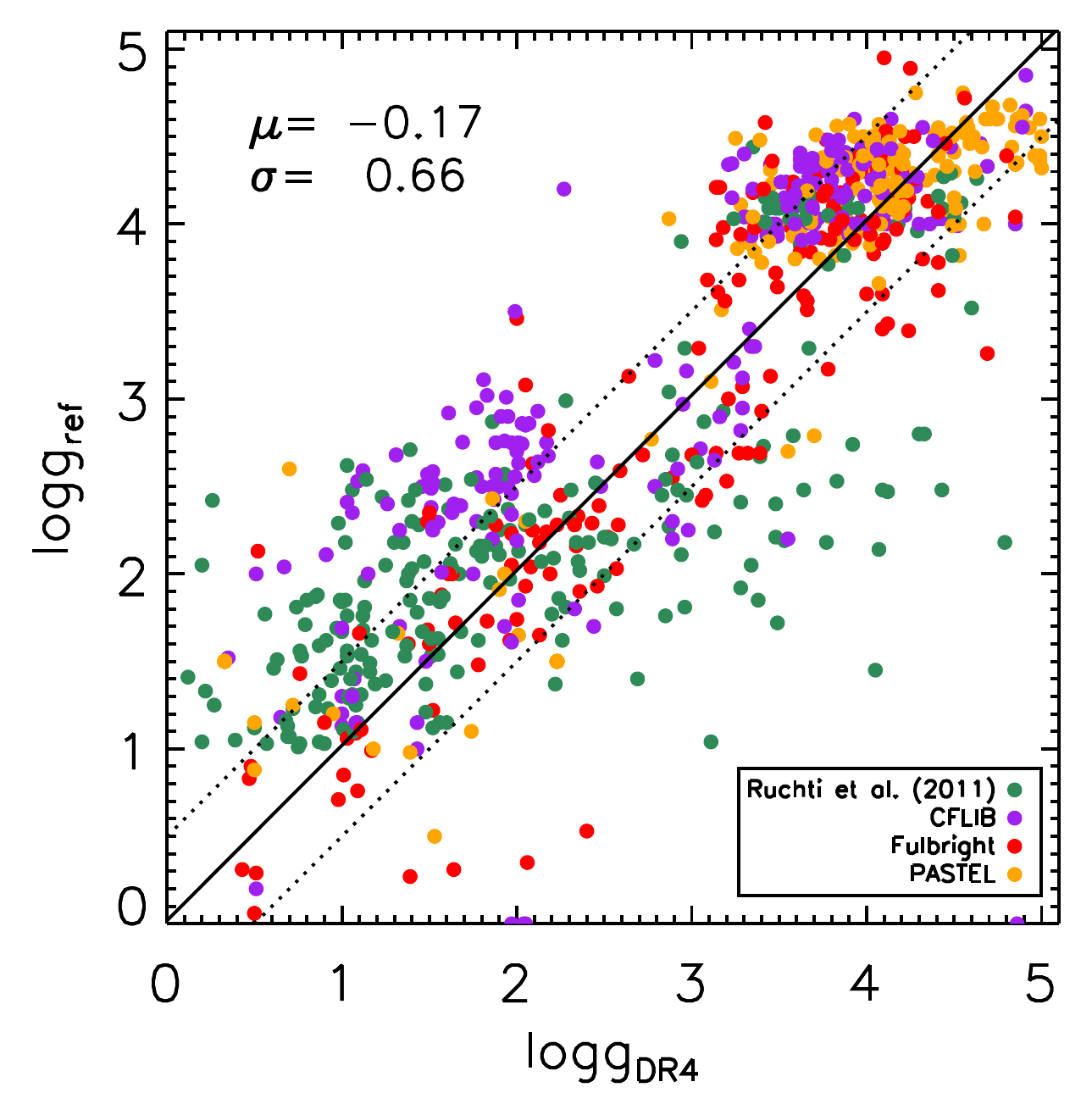}
\caption{{\bf Top left:} The histograms of the internal  radial velocity error for data new
  to each data release.  The bin size is 0.2~\kms.  For DR4, the number of
  stars per  bin is divided by 15  to compensate for the  increase in sample
  size. {\bf Bottom left:} The cumulative distributions, where the  dotted lines mark 50, 68 and
  95\% of the samples. {\bf Middle:} Comparison of the reference values found in the literature and the derived effective temperatures. Dotted diagonal lines represent offsets from unity of $\pm 300$~K. The mean offsets ($\mu$) and the dispersions ($\sigma$) are indicated in the upper left corner of each plot.
  {\bf Right:} Same for \logg, dotted diagonal lines represent offsets from unity of $\pm 0.5$~dex.}
\label{f:internal}
\end{figure}

Radial velocities (RV) are obtained using a standard cross-correlation in Fourier space on the continuum subtracted spectra (see DR3 for details) in a two-step procedure: First an estimate of the RV is determined using a subset of 10 template spectra. This first estimate gives an RV estimate with an accuracy better than 5 \kms\ and is used to shift the spectrum to the zero velocity frame. Using the full template database then a new template is constructed using a penalized chi-square technique (see DR2), which then in turn is used to derive a the more precise RV published in the data base. The histograms of the internal error of the RVs in DR1, DR2, DR3 and DR4 is shown in Fig.~\ref{f:internal}.  It can be seen that 68\% of the targets have an internal RV error better than 1.4 \kms\ in DR4 (see left frame of Figure 2). Error estimates derived from repeat observations and from comparison with external reference stars give values consistent with the numbers quoted above.

\subsection{Atmospheric parameters}

The wavelength region $\lambda \lambda 8410-8795$\AA~ is very attractive for Galactic archaeology purposes, as it features relatively few telluric absorptions, but exhibits many iron and $\alpha$-element lines, in particular the prominent Ca triplet, which allows relatively easily radial velocity measurements and metallicity estimations for any type of spectrum.  However, spectra with a resolution $R \le 10~000 $ suffer from spectral degeneracies that, if not appreciated,  can introduce serious biases in spectroscopic surveys that use automated parameterization pipelines.  These degeneracies are mostly important for cool main-sequence stars and stars along the giant branch (see Kordopatis \etal\ 2011 for a discussion). In matching observed spectra to templates in a library, decision-tree methods have been shown to perform better compared to other algorithms, like the projection methods (e.g.: principal component analysis) or the ones trying to solve an optimization problem (e.g.: minimum $\chi^2$), in particular when the SNR is low. The RAVE DR4 pipeline therefore employs a hybrid of  a decision-tree algorithm called DEGAS (Bijaoui \etal\ 2012) and a projection method called MATISSE (Recio-Blanco \etal\ 2006) which improves the interpolation between the grid points.  A comparison with a set of reference stars (see middle and right frame of Figure 2) shows a satisfactory dispersion of about 400K for the derived \teff\ (considerable less for a high S/N subsample), while the \logg\ determination is still suffering from the aforementioned degeneracies in the Ca triplet region (see table 1 and 2 in DR4).

\subsection{Chemical abundances}

Abundances for individual chemical elements are determined for the elements Mg, Al, Si, Ti, Ni and Fe based on a curve of growth analysis using the atmospheric parameters of the previous section as input values for \teff\ and \logg\ (Boeche \etal\ 2011). The chemical pipeline relies on an equivalent widths (EWs) library which contains the expected EWs of the lines visible in the RAVE wavelength range (604 atomic and molecule lines). These EWs are computed for a grid of stellar parameters values covering the range [4000,7000]~K in \teff, [0.0,5.0]~dex in \logg\ and $[-2.5,+0.5]$~dex in \meta\ and five levels of abundances in the range $[-0.4,+0.4]$~dex.  The chemical pipeline constructs on-the-fly spectrum models by adopting the effective temperatures and surface gravities obtained by the DR4 atmospheric parameters pipeline. It then searches for the best fitting model by minimizing the $\chi^2$ between the models and the observational data. The estimated errors in abundance, based on a comparison with reference stars, depend on the element and
range from 0.17~dex for Mg, Al and Ti to 0.3~dex for Ti and Ni. The error for
Fe is estimated as 0.23~dex.

\subsection{Distances}
Distances are estimated for spectra with  SNR$>10$~pixel$^{-1}$ by projecting the atmospheric parameters onto a the Padova set of theoretical isochrones and obtaining the most likely value of the absolute magnitude of the stars. DR4 employs an algorithm based on the Bayesian distance-finding method  presented in Binney \etal (2014), and takes into account the interstellar extinction, as well as kinematic correction factors obtained by the method of Sch\"onrich \etal\ (2012). Testing using Hipparcos stars indicates that the inverse of the expectation value of the parallax is the most reliable distance estimator. Compared to a subset of Hipparcos stars, the method results in an over-estimation of less than 10\% for the dwarfs and less than 20\% for the giants. The method has also been tested on the open cluster spectra, delivering very satisfactory distances, provided a cluster-specific age prior is used. 

\begin{figure*}
\centering

$\begin{array}{cc}
\includegraphics[width=0.47\linewidth, angle=0]{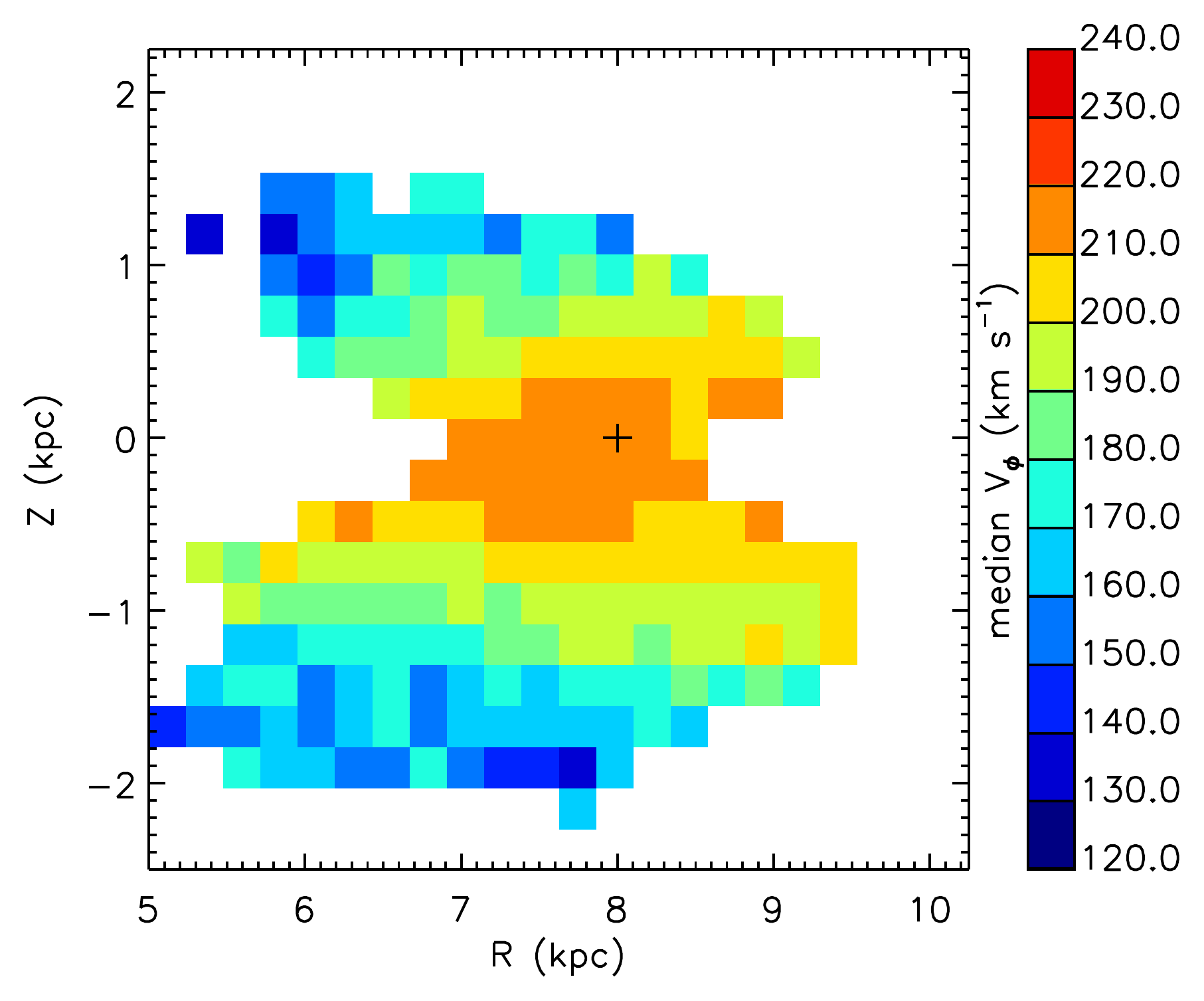}&
\includegraphics[width=0.47\linewidth, angle=0]{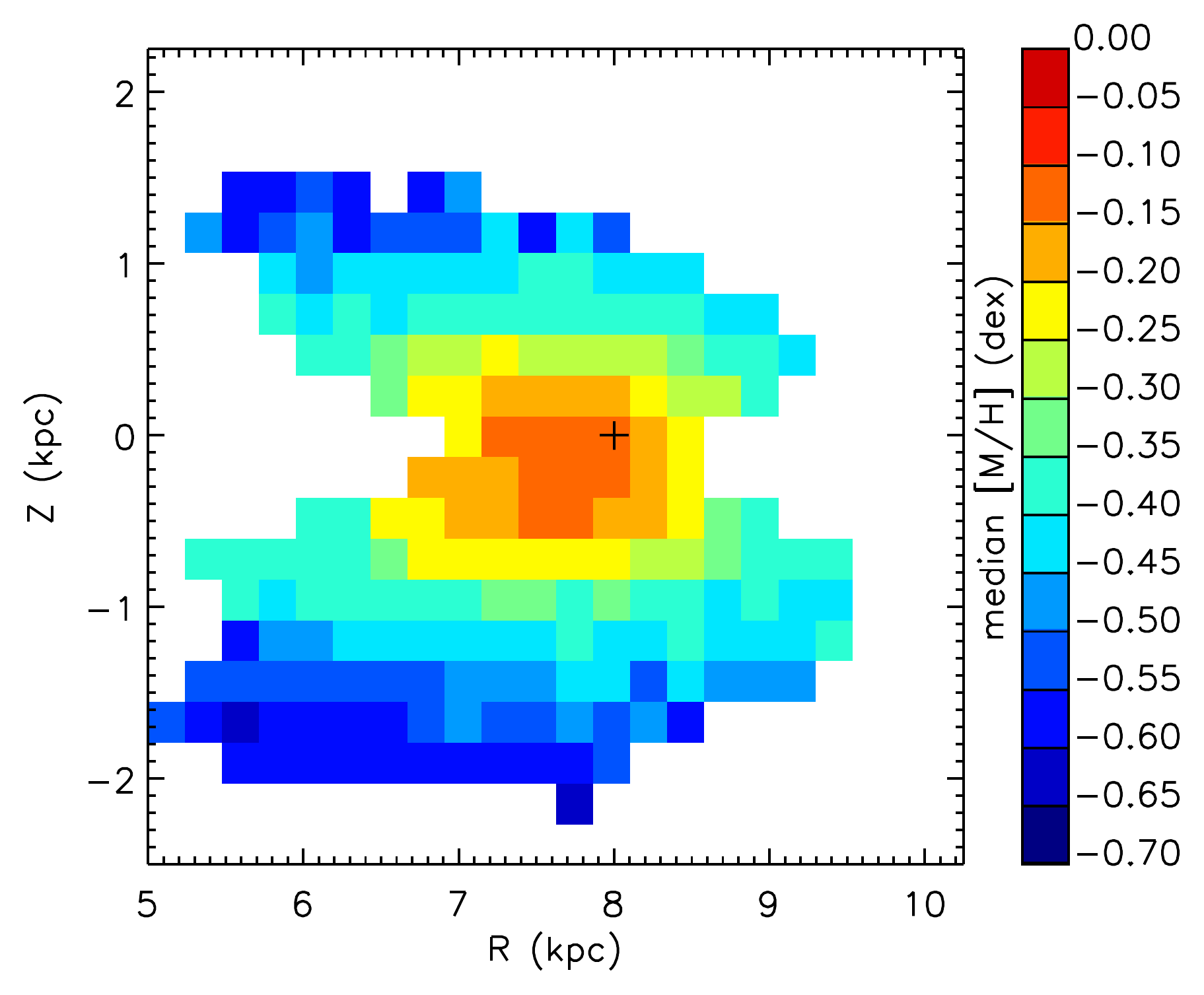}
\end{array}$

\caption{Median azimuthal velocities, $v_\phi$, and metallicities, $\rm [M/H]$, for all the RAVE stars for which distance and velocity determinations were available. The bins in $R$ and $Z$ are 0.25~kpc and contain at least 50 stars each. The black $"+"$ sign is at ($R_\odot=8$~kpc, $Z_\odot=0$~kpc), the assumed position of the Sun (from Kordopatis \etal\ 2013b).}
\label{fig:Overview}
\end{figure*}

\section{RAVE compared to the Gaia RVS}

Based on the currently available proper motions (UCAC4), radial velocities derived from RAVE spectra and distances derived from matching RAVE stellar parameters with isochrones, for 80\% of the stars in the RAVE volume, 3D velocities are known to better than 20 \kms. Combined with the large number of targets, this allows already detailed investigations on the structure and kinematics of the Milky Way disk(s) (see Figure 3). Recent examples include: the detection of a wave-like pattern in the stellar velocity distribution (Williams \etal\ 2013); an improved determination of the total mass of the Milky Way (Piffl \etal\ 2014a) and of the local dark matter density (Bienaym\'e \etal\ 2014, Piffl \etal\ 2014b); the identification of chemo-dynamical signatures with respect to a merger origin of the Galactic disk (Minchev \etal\ 2014); the identification of stars in the local neighborhood with supersolar metallicity, which likely have formed closer to the Galactic center and then moved outwards by radial migration (Kordopatis \etal\ 2015); the identification of stars tidally stripped of the globular clusters M22, NGC 1851 and NGC3201 (Kunder \etal\ 2014, Anguiano \etal\ 2015); the creation of pseudo-3D maps of the diffuse interstellar band at 8620\AA\ (Kos \etal\ 2014); and the detection of a metal-rich high-velocity star born in the Galactic disk (Hawkins \etal\ 2014).

The available astrometry of Milky Way stars will dramatically improve already with the first Gaia data releases (Prusti \etal\ 2012), bringing the accuracy of the 3D velocities down to better than 5 \kms, and at the same time considerably increasing the probed volume. Regarding the stellar parameters derived from Gaia spectroscopy, changes in terms of accuracy will be more gradual, as the wavelength range of Gaia-RVS and RAVE is almost identical, though the resolution of the Gaia-RVS is somewhat higher (10 500 for Gaia-RVS vs 7 500 for RAVE). However, the volume probed at similar accuracy will be 1-2 magnitudes deeper with much higher completeness, in particular at faint magnitudes. A particularly interesting data set for the next few years will be the combination of Gaia astrometry, as published in the early data releases, with RAVE stellar parameters, radial velocities and abundances, as according to the current Gaia release schedule, the latter properties will only be provided in the later data releases.

\section{Outlook}

The RAVE collaboration currently investigates a number of improvements for the next data release. Besides analyzing the as of yet not published stars, a recalibration for super solar metallicities has been devised (Kordopatis \etal\ 2015).  The distance pipeline has been found to systematically underestimate the distances for low metallicities ([Fe/H]$< -1$), as it is e.g. for stars stripped off globular clusters. A correction using lower-metallicity isochrones is underway. The release of optical photometry of APASS (Munari \etal\ 2014) will allow to determine \teff\ with the aid of additional photometric priors, reducing the degeneracy in the \logg\ determination. Last but not least, we expect to get additional calibration data for \logg\ from Kepler-2 astroseismology. Kepler-2 campaigns 0-2 currently have some 1400 RAVE stars on their target list.\\

{\footnotesize
Funding for RAVE has been provided by: the Australian Astronomical Observatory; the Leibniz-Institut 
f\"ur Astrophysik Potsdam (AIP); the Australian National University; the Australian Research Council; the French National Research Agency; the German Research Foundation (SPP 1177 and SFB 881); the European Research Council (ERC-StG 240271 Galactica); the Instituto Nazionale di Astrofisica at Padova; The Johns Hopkins University; the National Science Foundation of the USA (AST-0908326); the W. M. Keck foundation; the Macquarie University; the Netherlands Research School for Astronomy; the Natural Sciences and Engineering Research Council of Canada; the Slovenian Research Agency; the Swiss National Science Foundation; the Science \& Technology Facilities Council of the UK; Opticon; Strasbourg Observatory; and the Universities of Groningen, Heidelberg and Sydney. The research leading to these results has received funding from the European Research Council under the European Union's Seventh Framework Programme (FP7/2007-2013)/ERC grant agreement no. 321067. The RAVE web site is at http://www.rave-survey.org.
}

%%-----------------------------
%%      your bibliography
%%-----------------------------


\begin{thebibliography}{99}
\bibitem[2015]{Anguiano2015}Anguiano, B., \etal\ 2015, MNRAS, submitted
\bibitem[2014]{Bienayme} Bienaym\'e, O. \etal\ 2014, A\&A 571, 92
\bibitem[2011]{degas}Bijaoui, A., Recio-Blanco, A., de Laverny, P., \& Ordenovic, C. 2012, Statistical methodology, 9, 55
\bibitem[2011]{Binney2014} Binney, J., \etal\ 2014, MNRAS 437, 351
\bibitem[2011]{B11} Boeche, C., \etal\ 2011, AJ 142, 193 (B11)
\bibitem[2014]{Hawkins2014} Hawkins, K., \etal\ 2014, MNRAS 447, 2046 
\bibitem[2011]{Kordopatis2011} Kordopatis, G., \etal\ 2011, A\&A 535, A106
\bibitem[2013a]{dr4} Kordopatis, G., \etal\ 2013, AJ 146, 134 (DR4)
\bibitem[2013b]{Kordopatis2013b} Kordopatis, G., \etal\ 2013, MNRAS 436, 3231 
\bibitem[2015]{Kordopatis2015} Kordopatis, G., \etal\ 2015, MNRAS 447, 3526 
\bibitem[2015]{Kos2014} Kos, J., \etal\ 2014, Science 345, 791 
\bibitem[2014]{Kunder2014} Kunder, A., \etal\ 2014, A\&A 572, 30 
\bibitem[2014]{Minchev2014} Michev, I., \etal\ 2014, ApJL 781, L20 
\bibitem[2014]{APASS}Munari, U., \etal\ 2014, AJ 148, 81
\bibitem[2004]{GCS}Nordstr\"om, B., \etal\ 2004, A\&A 418, 989
\bibitem[2014a]{Piffl2014a} Piffl, T., \etal\ 2014, A\&A 562, 91 
\bibitem[2014b]{Piffl2014b} Piffl, T., \etal\ 2014, MNRAS 445, 3133 
\bibitem[2012]{Prusti2012} Prusti, T., \etal\ 2012, AN 333, 453 
\bibitem[2006]{matisse}Recio-Blanco, A., Bijaoui, A., \& de Laverny, P. 2006, MNRAS
370, 141
\bibitem[2012]{Schoenrich2012}Sch\"onrich, R., Binney, J., \& Asplund, M. 2012, MNRAS 420,
1281
\bibitem[2011]{dr3} Siebert, A., \etal\ 2011, AJ 141, 187 (DR3)
\bibitem[2006]{dr1} Steinmetz, M., \etal\ 2006, AJ 132, 1645 (DR1)
\bibitem[2013]{Williams2013} Williams, M., \etal\ 2013, MNRAS 436, 101 
\bibitem[2008]{dr2} Zwitter, T., \etal\ 2008, AJ 136, 421 (DR2)
\end{thebibliography}
\end{document}